\documentclass[10pt, conference, letterpaper]{IEEEtran}
\pdfoutput=1 
\IEEEoverridecommandlockouts
\usepackage{cite}%[nocompress]{cite}
\usepackage{amsmath,amssymb,amsfonts}
\usepackage{comment}
\usepackage{algorithmic}
\usepackage{booktabs}
\usepackage{graphicx}
\usepackage{textcomp}
\usepackage{xcolor}
\usepackage[pscoord]{eso-pic}
\usepackage{balance}
\usepackage{fancyvrb}

\usepackage{algorithm2e}
\RestyleAlgo{ruled}
\SetKwComment{Comment}{/* }{ */}

\def\BibTeX{{\rm B\kern-.05em{\sc i\kern-.025em b}\kern-.08em
    T\kern-.1667em\lower.7ex\hbox{E}\kern-.125emX}}
    
\newcommand{\placetextbox}[3]{% \placetextbox{<horizontal pos>}{<vertical pos>}{<stuff>}
  \setbox0=\hbox{#3}% Put <stuff> in a box
  \AddToShipoutPictureFG*{% Add <stuff> to current page foreground
    \put(\LenToUnit{#1\paperwidth},\LenToUnit{#2\paperheight}){\vtop{{\null}\makebox[0pt][c]{#3}}}%
  }%
}%
\def\BibTeX{{\rm B\kern-.05em{\sc i\kern-.025em b}\kern-.08em
    T\kern-.1667em\lower.7ex\hbox{E}\kern-.125emX}}
\def\BibTeX{{\rm B\kern-.05em{\sc i\kern-.025em b}\kern-.08em
    T\kern-.1667em\lower.7ex\hbox{E}\kern-.125emX}}

\placetextbox{0.5}{0.96}{\texttt{\textcolor{red}{\textbf{Accepted for presentation at ISQED'25 - DOI/link to IEEE Xplore will be updated}}}}

\usepackage[frozencache,cachedir=.]{minted} 

\usepackage{listings}

\usepackage{pifont}
\usepackage{framed}
\usepackage{soul}
\usepackage{booktabs}  
\usepackage{multirow}
\usepackage{acronym}
\acrodef{IP}[IP]{intellectual property block}
\acrodef{SoC}[SoC]{System-on-Chip}
\acrodef{IC}[IC]{integrated circuit}
\acrodef{eFPGA}[eFPGA]{embedded field programmable gate array}
\acrodef{RTL}[RTL]{register-transfer level}
\acrodef{CPS}{Cyber-Physical System}
\acrodef{IoT}{Internet of Things}
\acrodef{CAD}{Computer-Aided Design}
\acrodef{EDA}{Electronic Design Automation}
\acrodef{HPC}{High-Performance Computing}
\acrodef{DL}{deep learning}
\acrodef{ML}{machine learning}
\acrodef{NLP}{natural language processing}
\acrodef{IC}{Integrated Circuit}
\acrodef{CWE}[CWE]{Common Weakness Enumeration}
\acrodef{CVE}[CVE]{Common Vulnerabilities and Exposures}
\acrodef{LLM}[LLM]{large language model}
\acrodef{NMT}[NMT]{neural machine translation}
\acrodef{IP}[IP]{hardware intellectual property block}
\acrodef{HDL}[HDL]{hardware description language}
\acrodef{RTL}[RTL]{register-transfer level}
\acrodef{SDL}[SDL]{security development lifecycle}
\acrodef{FSM}[FSM]{finite state machine}
\acrodef{AST}[AST]{abstract syntax tree}
\acrodef{SoC}[SoC]{system-on-chip}
\acrodef{SA-EDI}{Security Annotation for Electronic Design Integration standard}
\usepackage{array}
\newcolumntype{L}[1]{>{\raggedright\let\newline\\\arraybackslash\hspace{0pt}}m{#1}}
\newcolumntype{C}[1]{>{\centering\let\newline\\\arraybackslash\hspace{0pt}}m{#1}}
\newcolumntype{R}[1]{>{\raggedleft\let\newline\\\arraybackslash\hspace{0pt}}m{#1}}
\usepackage{url}
\usepackage[caption=false,font=footnotesize,labelformat=simple]{subfig}

\usepackage[bookmarks=false,hidelinks]{hyperref}

\newcommand{\rt}[1]{{\color{red}{#1}}}

\begin{document}
\bstctlcite{IEEEexample:BSTcontrol}

\title{%
    Toward Automated Potential Primary Asset \\Identification in Verilog Designs%
}

\author{%

    \IEEEauthorblockN{Subroto Kumer Deb Nath and Benjamin Tan}
    \IEEEauthorblockA{\textit{Department of Electrical and Software Engineering} \\
    \textit{Schulich School of Engineering} \\
    \textit{University of Calgary}\\
    \{subroto.nath,benjamin.tan1\}@ucalgary.ca}
    \thanks{
    
    This research is supported in part by Alberta Innovates, and the Natural Sciences and Engineering Research Council of Canada (NSERC) [RGPIN-2022-03027]. Cette recherche a été financée en partie par le Conseil de recherches en sciences naturelles et en génie du Canada (CRSNG).
    This research work is partly supported by a gift from Intel Corporation. This work does not in any way constitute an Intel endorsement of a product/supplier. 

    \rt{\textcopyright 2025 IEEE. Personal use of this material is permitted. Permission from IEEE must be obtained for all other uses, in any current or future media, including reprinting/republishing this material for advertising or promotional purposes, creating new collective works, for resale or redistribution to servers or lists, or reuse of any copyrighted component of this work in other works
    }}
}

\IEEEtitleabstractindextext{
\begin{abstract}
With greater design complexity, the challenge to anticipate and mitigate security issues provides more responsibility for the designer. 
As hardware provides the foundation of a secure system, we need tools and techniques that support engineers to improve trust and help them address security concerns. 
Knowing the security assets in a design is fundamental to downstream security analyses, such as threat modeling, weakness identification, and verification. 
This paper proposes an automated approach for the initial identification of potential security assets in a Verilog design.
Taking inspiration from manual asset identification methodologies, we analyze open-source hardware designs in three IP families and identify patterns and commonalities likely to indicate structural assets. 
Through iterative refinement, we provide a potential set of primary security assets and thus help to reduce the manual search space. 
\end{abstract}
\begin{IEEEkeywords}
    Asset Identification, Verilog, Automation, Hardware Security
\end{IEEEkeywords}
}

\maketitle
\IEEEdisplaynontitleabstractindextext

\section{Introduction}

The development of secure \acp{SoC} requires diligent consideration of security concern through a security development lifecycle (SDL)~\cite{Khattri_HSDL_2012}. 
To address security issues, design and verification teams should identify potential \textit{assets} in a system to make decisions regarding their protection and to help inform verification of any security features. 
Ideally, designers should be proactive in identifying and mitigating potential security threats, ultimately leading to more secure and trustworthy hardware systems.

Given that the detection and mitigation of potential security threats in the pre-silicon stage is ideal (because early changes are considerably easy and more cost-effective~\cite{Ahmad_2022}), designers require more tools that can help with security analysis in earlier stages of the design flow. 
This analysis starts by identifying primary security assets that need robust protection against security threats and attacks~\cite{mishra_hardware_2017}, currently a predominantly\textbf{ manual and subjective} process that relies on some level of security expertise~\cite{ieee_p3164_working_group_asset_2024}. Prior work has suggested that while some ``primary assets are often seen as `self-evident''' others are ``not \ldots{} immediately obvious''~\cite{Ayalasomayajula_Automatic_2024}. 
This initial step is crucial as it sets the direction for subsequent security measures and analyses.
Moreover, the evolving hardware security landscape has led to initiatives such as Accellera's \ac{SA-EDI}~\cite{accellera} and the related IEEE P3164 effort to establish a standard for \acp{IP} security collateral; these start with \textit{identifying assets}. 

Thus, in this work, we investigate whether it is possible to \textit{automate} initial potential \textbf{primary} asset identification. 
\textbf{This is in contrast to prior work that focuses on secondary asset identification} only~\cite{farzana_saif_2021,Ayalasomayajula_Automatic_2024}. 
Our proposed approach is based on identifying patterns in designs, and we explore a proof-of-concept by studying a series of open-source designs of several IP family types. 
Designers can adapt the proposed process to identify assets in their own projects. 
By reducing the manual workload faced by designers, this approach aims to streamline the early stages of security analysis. 
Our contributions are as follows:
\begin{itemize}
    \item Analysis of open source hardware designs for guidance on patterns related to assets in \ac{RTL} source code written in Verilog/SystemVerilog.
    \item A proposed approach for automating the primary asset identification that uses insights from our analyses of open-source hardware.
    \item Experimental evaluation of the proposed approach showing a True Positive Rate of 82.18\% compared to manually labeled asset lists.
\end{itemize}

\section{Background and Related Work\label{sec:background}}

\begin{figure*}[t]
\centering
\subfloat[Crypto]{\label{fig:Crypto-Graph}\includegraphics[width=0.33\textwidth]{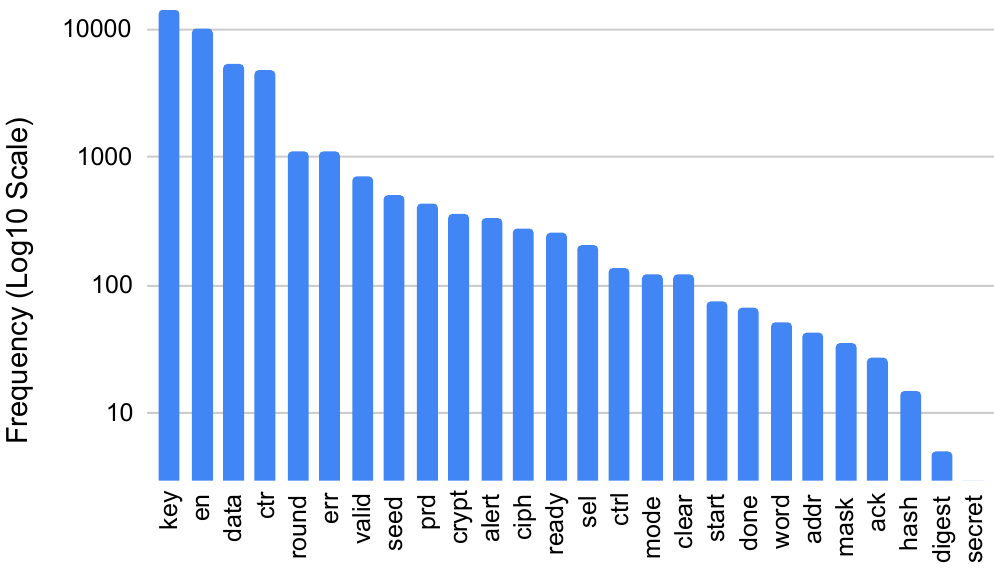}}\hfill
\subfloat[Interface-GPIO]{\label{fig:GPIO-Graph}\includegraphics[width=0.33\textwidth]{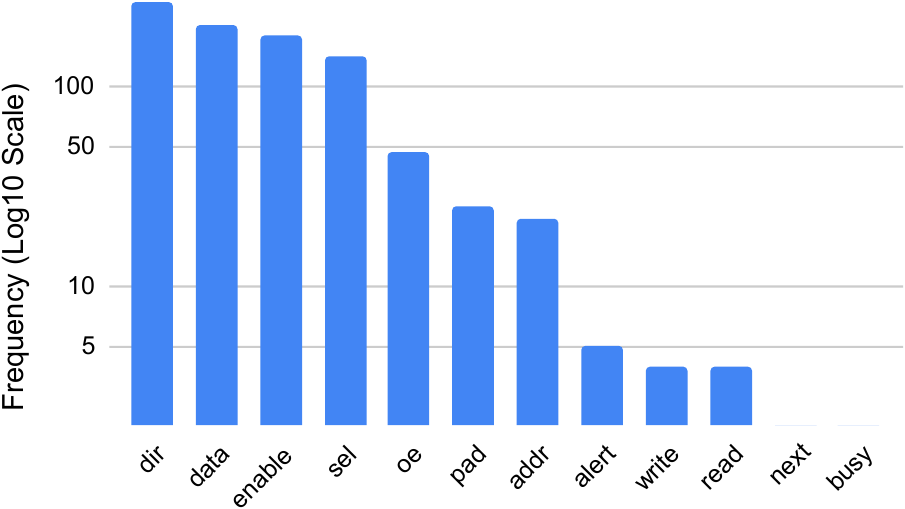}}\hfill
\subfloat[Interface-Peripheral]{\label{fig:Peripheral-Graph}\includegraphics[width=0.33\textwidth]{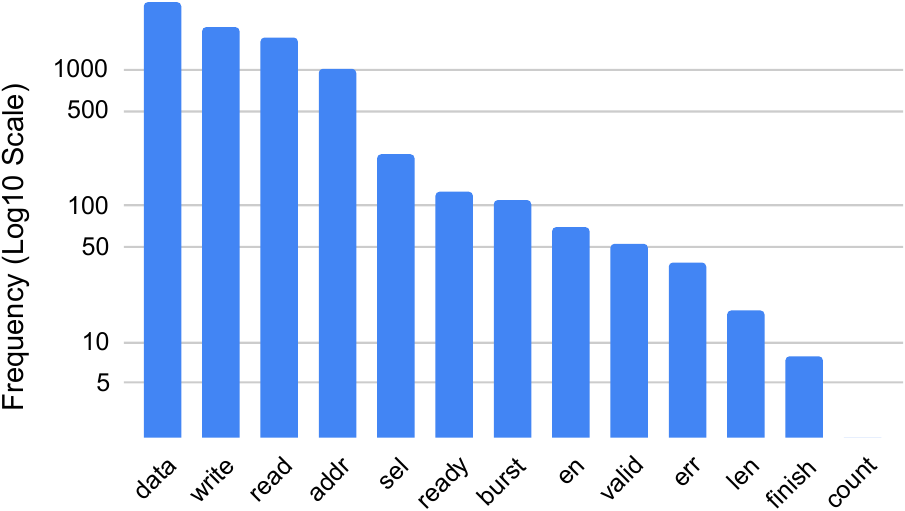}}
\caption{The number of occurrences for our identified partial keywords for three IP families.\label{fig:explanations_of_general_image}
}
\end{figure*}

An asset is any physical or logical component of \textit{value} and is essential for the proper functioning or security of the system\cite{holdings2009arm}.
The elements in an \ac{IP} that process, control, and store important values and interact with other \acp{IP} in an \ac{SoC} and communicate with the external peripherals are considered as \textit{primary} assets. 
Secondary assets are mostly internal design components of an \ac{IP} that help to propagate and handle the primary assets throughout the entire \ac{IP}.
It is helpful to know the assets in the system for several purposes, such as formulating security properties or identifying potential attack points (e.g., as in Accellera's \ac{SA-EDI} standard~\cite{accellera}). 
Recently, an IEEE P3164 white paper proposed a Conceptual and Structural Asset (CSA) methodology to help manually identify primary assets~\cite{ieee_p3164_working_group_asset_2024}, especially considering security objectives of \textbf{confidentiality}, \textbf{integrity}, and \textbf{availability} (the ``CIA triad'')~\cite{noauthor_what_nodate}, as well as the risk for ``undermined expected behavior'' in normal operation. 
Additionally, initiatives like MITRE's \acp{CWE}~\cite{mitreCommonWeakness} offer various examples of hardware weakness for identifying and mitigating vulnerabilities in hardware systems. 
This standard and methodologies, along with \ac{CWE} examples, give us valuable insights into current security challenges and how to potentially avoid them. 

Even so, as observed by the authors in~\cite{10140100}, engineers face challenges in identifying assets in the initial stages of hardware verification. 
Engineers must assess asset weaknesses, including proper initialization, information flow, and access controls. This process demands deep knowledge of security assets~\cite{ray_system--chip_2018}. However, there is a lack of tools to aid and automate this task, making it even more challenging for engineers.
Prior work attempted to perform generalized security analysis of \ac{HDL} code~\cite{Ahmad_2022} using syntactical patterns but found that more information from a designer (such as assets) was needed to address false positives. 
The authors of~\cite{polian_introduction_2017} described two situations where an element or the \ac{IP} itself as a whole can be an asset.
The authors of~\cite{meza_security_2023} demonstrated the security properties verification method with the help of security assets, where identifying hardware assets is the most crucial (manual) part. 
The authors of~\cite{farzana_saif_2021, Ayalasomayajula_Automatic_2024} proposed an automatic \textit{secondary} asset detection algorithm, that assumes that \textit{manual} \textbf{primary} security asset identification was performed. 
To the best of our knowledge, no prior work currently exists for \textbf{automated \underline{primary} asset identification} in Verilog source code, so our work complements prior works by offering an automated approach to identify potential primary assets. 

\section{Proposed Approach\label{sec:proposed}}

\subsection{Insights from Existing Designs\label{ssec:IED}}
We first discuss some insights from analyzing some existing open-source designs. 
We investigated 22 open-source cryptography IPs, 13 Peripheral Interface IPs, and 12 GPIO IPs (a subset of the family types that are defined in~\cite{accellera}) to gain insights into their design characteristics and commonalities, with a particular focus on 
recurring textual features that can be leveraged to develop an automated potential primary asset detection tool. 
These designs were collected from several sources\footnote{%
To support the community's efforts in this work, we make the details of our identified asset list and the associated IPs available here: \url{https://github.com/CalgaryISH/Asset_Dataset_using_PKG}
}, including GitHub, OpenCores, and OpenTitan, some of which have been used in prior work (e.g.,~\cite{Ahmad_2022}).
We manually examined design documentation and source code to get an intuitive sense of commonalities across IPs in each family. 

We applied the manual CSA method~\cite{ieee_p3164_working_group_asset_2024} to identify potential security assets. We also observed that most open-source designs involved ``sensible'' names for elements in the design (e.g., signals, sub-modules), and that frequently occurring signal/variable names in a given IP family provide indicators of ``important'' elements that point to assets. 

From this manual analysis of different \ac{IP} families and their source code, we curated a list of ``partial keywords'' for each \ac{IP} family that was associated with security-relevant Inputs, Outputs, and Reg/Wire/Logic(Net) signals, and we excluded ``typical'' names or tokens that were usually irrelevant to security, to increase the robustness of ``important RTL elements'' detection in the matching stage of our tool. 
This extends the similar concept of ``keyword-matching'' described in prior work~\cite{Ahmad_2022}.

For example, ``en'' represents the ``Enable'' partial keyword group ($PKG_{enable}$), which can be usable to detect signals named ``write\_en'', ``write\_enable'', ``wen'', ``gpio\_oen'', ``gpio\_out\_ena'' etc.
We sometimes added ``partial keywords'' that were especially meaningful for a given IP family even though they were not the most frequently appearing based on our manual code analysis. 
For instance, we added a partial keyword ``text'' in the list for Crypto to detect signals like ``plain\_text'', ``text\_in'', and ``text\_out'', which are important but not very common. 
\autoref{fig:explanations_of_general_image} illustrates the frequency with which the elements of our list of ``partial keywords'' appear throughout all files in an IP class. 
We also considered the following factors in the curation of partial keywords.

\subsubsection{Location and Frequency}
The location of the signals is crucial to understanding their nature. Potential important signal names often appear in logical/conditional expressions, assignments in sequential blocks, and combinational blocks. 
The frequency depends on the file size, but frequent signal names should be considered for further processing. These can be potential candidates if they frequently appear in these aforementioned locations.

\subsubsection{Keywords and Tokens of the Language}
Every language has reserved keywords, tokens, and constant names, which are special identifiers reserved to define the language constructs (e.g., \textit{wire}). We ignored everything that completely matched with these special ``words''.

\subsubsection{Types and Signal Width} We considered Input, Output, and internal Net-type signals in this study. Also, for signal width, we considered three categories: 1-bit, 2-bit to 8-bit, and larger signals up to 256-bit. A partial keyword group with a similar signal type(Input, Output, Net) and width is more desirable as it reduces the rule formulation complexities for asset classification.

\subsubsection{Overlap in Names} Due to naming conventions and functionalities of the designs, we observed some partial and full overlap in the signal names for different projects in the same IP class. As we discussed before, we used this overlapping portion of the name to create a partial keyword group like ($PKG_{enable}$). However, signal type and width should be taken into consideration as well.

\subsubsection{Commonalities in Roles} In some cases, we could not find any similarities in signal names. In that case, we created a partial keyword group according to the signal type, width, and role in the design. In this case, we used our design knowledge and ideas from the behavioral patterns. A partial keyword group for the status signal can be a good example. We have mentioned the most common status signals in \autoref{sssec:status}.

\subsection{Behavioral Patterns\label{ssec:BP}}

As simple name matching is insufficient to identify potential assets, we also manually examined how the potential candidate signals identified ``behave'' and ``function'' inside a design (e.g., if they often appear in \texttt{if-else}, \texttt{always} or \texttt{assign} statements, what are the ``width'' of the signals, in which operations they are associated to) to classify a set of signals' behavior patterns. 
Our approach does not solely depend on the keywords. 
Keywords help to identify potential candidate signals for the behavioral pattern detection stage. The behavioral pattern, including the attributes, functionalities, and location of a signal, indicates the overall structure, which is crucial to detecting structural assets~\cite{ieee_p3164_working_group_asset_2024}.

We observed four common signal behavioral patterns in the open-source hardware designs. 
Based on our manual analysis, attributes of the signals, and how they function in a design, we developed an algorithm denoted by \autoref{alg:one} to detect all four types of signals. 
\autoref{fig:code-example} provides an example of Verilog code that we use next as a running example to explain behavioral patterns. 
Each type of signal behavior can be described as relevant to confidentiality, availability, and/or integrity security properties. 

\subsubsection{Control Signals} are typically single-bit wide input signals or nets/variables (assigned or instantiated with input port in the module) that appear inside the conditional expression of \texttt{if-else} blocks, \texttt{case} blocks, and \texttt{ternary operation} statements. 
They are responsible for enabling/disabling functionality, controlling data flows and value assignment, and are often used to clear or load information from or into a memory register. Control signals are responsible for managing and controlling one or more blocking or non-blocking assignment statements inside \texttt{if-else} block, \texttt{case} block, and \texttt{ternary operation}. 
A control signal of a module can be connected with a status signal from a different module through instantiation when an interdependent sequential process occurs between two modules.
The \textit{Availability} security objective is commonly relevant to the Control signals. 
In the example (\autoref{fig:code-example}), \texttt{load} is a 1-bit input signal and controls the 128-bit data loading to \texttt{data\_in\_reg} register on line number 15. If the \texttt{load} becomes 0, no data will be loaded to the \texttt{data\_in\_reg} register.

\subsubsection{Configuration Signals} are typically 2-bit to a few bits wide (in most cases, up to 8-bit) input signals or nets/variables (assigned or instantiated with input port in the module) that appear mostly in \texttt{case} expression, \texttt{ternary operation}, and conditional expression in multi-statements containing \texttt{if-else} blocks. 
These signals configure the operational flow of a module, data read and write direction, data splitting and loading into multiple memory registers or clear from the memory registers, select multiplexer's outputs, and control state transitions. 
\textit{Availability} and \textit{Integrity} security objectives are usually associated with the Configuration signals.
In the example, the 2-bit \texttt{bank\_selector} is a Configuration signal. According to code line 18, \texttt{bank\_selector} works as a Multiplexer output selector in the \texttt{case} expression for loading the split data into four memory banks.

\begin{figure}
\centering
\begin{minipage}{0.9\columnwidth}
\begin{minted}[breaklines=true,fontsize=\scriptsize, linenos=true]{verilog}
module data_splitter (
    input clk,
    input load,
    input [1:0] bank_selector,
    input [127:0] data,
    output reg [31:0] bank0,
    output reg [31:0] bank1,
    output reg [31:0] bank2,
    output reg [31:0] bank3,
    output reg done
);
  reg [127:0] data_in_reg;
  reg done0, done1, done2, done3;
  always @(posedge clk) begin
    if (load) data_in_reg <= data;
  end
  always @(data_in_reg or bank_selector) begin
    case (bank_selector)
      2'b00: begin
        bank0 <= data_in_reg[31:0];
        done0 <= 1'b1;
      end
      2'b01: begin
        bank1 <= data_in_reg[63:32];
        done1 <= 1'b1;
      end
      2'b10: begin
        bank2 <= data_in_reg[95:64];
        done2 <= 1'b1;
      end
      2'b11: begin
        bank3 <= data_in_reg[127:96];
        done3 <= 1'b1;
      end
      default: begin
      end
    endcase
  end
  always @(posedge clk) begin
    if (done0 && done1 && done2 && done3) 
        done <= 1'b1;
    else done <= 1'b0;
  end
endmodule
\end{minted}
\end{minipage}

\caption{A simple 128-bit to four banks of 32-bit data splitter code example.}
\label{fig:code-example}

\end{figure}

\begin{algorithm}[t!]
\footnotesize

\caption{Algorithm to detect different behavioral patterns in signals}\label{alg:one}

\KwData{RTL Code Written in Verilog or SystemVerilog, input\_ports[\ ], output\_ports[\ ], net\_variables[\ ];}
\KwResult{Control\_Signals[\ ], Configuration\_Signals[\ ], Status\_Signals[\ ], Data\_Signals[\ ];}
\vspace{1mm}
\For{$line\leftarrow 1$ \KwTo max\_line\_number of RTL\_Code} {

\If{$line\ contains\ a\ conditional\ expression$} {
\tcc{$x$ is the signal name inside the conditional expression}
\If{Width of $x ==\ 1bit$}{

\If{$x \in input\_ports[\ ]\ ||\ net\_variables[\ ] $} {append $x$ to Control\_Signals[\ ];}

}
\ElseIf{Width of $x \geq 2bit\ \&\&$ The conditional block contains multiple statements }{

\If{$x \in input\_ports[\ ]\ ||\ net\_variables[\ ] $} {append $x$ to Configuration\_Signals[\ ];}

}

}

\ElseIf{$line\ contains\ a\ blocking\ or\ non-blocking\ assignment$}{
\tcc{$l$ and $r$ are the signal names that appear on the left-hand side and right-hand side of an assignment statement, respectively}
\If{Width of $l ==\ 1bit\ \&\&\ l \in output\_ports[\ ] $}{ 
{append $l$ to Status\_Signals[\ ];}
}
\ElseIf{Width of $l \geq 2bit\ \&\&\ l \in output\_ports[\ ]$}{append $l$ to Data\_Signals[\ ];}
\ElseIf{Width of $r \geq 2bit\ \&\&\ r \in input\_ports[\ ]$}{append $r$ to Data\_Signals[\ ];}

}
\Else{do nothing;} 
}

\end{algorithm}

\subsubsection{Status Signals\label{sssec:status}} are typically single bit-width output signals or nets/variables (assigned or instantiated with output port in the module) that appear on the left-hand side of blocking/non-blocking assignments, statements under conditional code segments (e.g., \texttt{if-else} blocks, \texttt{case} blocks, and \texttt{ternary operation}). 
Usually, a status signal lets other modules know the status of a process or operation from the module to which it belongs. 
Common Status signals include \texttt{finish}, \texttt{done}, \texttt{ready}, \texttt{success}, \texttt{alert}, and \texttt{error}. 
If a status signal is connected with a control signal of another module, the \textit{Availability} objective is relevant; otherwise, \textit{Integrity} security objective is commonly relevant to the Status signals.
In the example, \texttt{done} is a Status signal. From code lines 40 to 42, the \texttt{if-else} block assigns the value of \texttt{done} signal based on \texttt{done0}, \texttt{done1}, \texttt{done2}, and \texttt{done3} signals. 

\subsubsection{Data Signals} can be inputs or outputs with a multi-bit width for assigning storage addresses, memory registers, information for processing, and processed information in a module.
Data signals propagate data and are processed in multiple modules in an \ac{IP}.
There is another kind of Data signal that does not get changed or processed during an operation but is involved in security-critical operations like encryption and decryption. ``Seed'' and ``Key'' are typical examples of this Data signal. 
Data signals appear in both blocking and non-blocking assignment-type statements in a module.
Data signals containing critical information are directly linked to \textit{Confidentiality}.
In our example, the 128-bit \texttt{data} and 32-bit \texttt{bank0}, \texttt{bank1}, \texttt{bank2}, and \texttt{bank3} are responsible for storing and carrying information; these all Data-type behavioral pattern.
The \texttt{data} may contain sensitive information like \textit{key} for a key-expansion operation in an encryption or decryption IP, or it may contain plain user information that needs to be protected equally. 
In line 15 of the \autoref{fig:code-example}, \texttt{data} appears on the right-hand side of the non-blocking assignment.
Conversely, all four banks appear on the left-hand side of the non-blocking assignment on lines 20, 24, 28, and 32, respectively. 
Such information, like which side of the assignment the signal appears, can be useful for automated asset identification. 

\begin{figure*}[t]
    \centering
    \includegraphics[width=\linewidth]{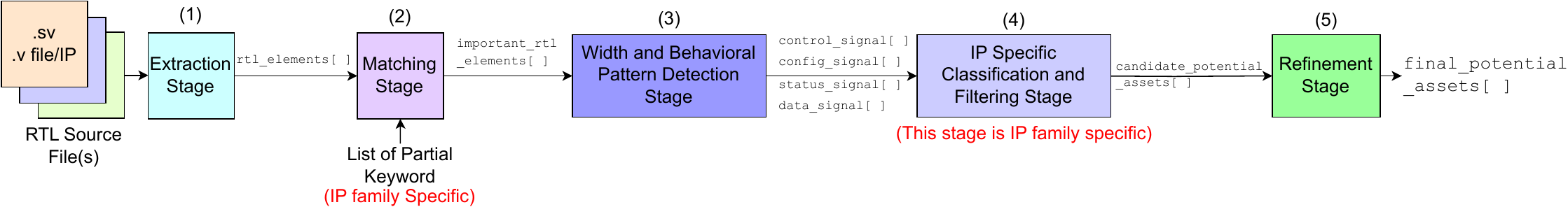}
    \caption{Overall view of our proposed automatic potential asset detection algorithm.}
    \label{fig:overview-figure}
\end{figure*}

\subsection{Automation}
In this section, we build on previously discussed observations and explain our proposed five-stage process for automated potential primary asset identification, as illustrated in \autoref{fig:overview-figure}. 

\subsubsection{Extraction Stage}
This stage uses RTL source files in Verilog or SystemVerilog. The tool iterates through all the available files in an \ac{IP}'s directory, extracting all the Input and Output Ports and Wire/Reg/Logic Nets. The tool stores these extracted ``RTL Elements'' in lists for the next stage. 

\subsubsection{Matching Stage}
In this stage, the tool finds partial matches between the RTL Elements and the IP-family-specific partial keyword lists to detect important signals that have the most potential to be assets. 
This stage outputs the subset of elements that are likely to be important from a security-perspective. 
Further pruning in the next stages narrows the potential candidate asset list. 
After this stage, the tool prepares a list of important\_rtl\_elements[ ] (reduced version of rtl\_elements[ ]) for the next stage.

\subsubsection{Width and Behavioral Pattern Detection Stage}
The matching step can sometimes result in 80\% of the RTL elements matching due to naming and multiple matches with partial keywords. 
For example, a signal \texttt{key\_rounding\_enable} in a Cryptographic \ac{IP} has three partial matches with keywords \texttt{en}, \texttt{round}, and \texttt{key}. 
The tool will detect \texttt{key\_rounding\_enable} for three partial keywords as a candidate for Important RTL Elements. 
However, not all of these should be considered structural assets (as explained in the IEEE P3164 white paper~\cite{ieee_p3164_working_group_asset_2024}). 
For this particular example, if the signal's width is single-bit and it is used to control any functionality or operation in the design (i.e., it has a ``control signal'' behavioral pattern), it can be a potential asset. 
Again, if the signal does not have the control pattern, just the partial keyword matching, it might not be a potential asset.
Therefore, we add another step to detect the signal width and behavioral patterns, as we previously described in \autoref{ssec:BP}. This will help the tool create lists of control\_signal[ ], config\_signal[ ], status\_signal[ ], and data\_signal[ ]. 

\subsubsection{IP-Specific Asset Classification and Filtering Stage}
Based on our manual analysis of the open-source designs, we designed a set of IP-family-specific classification rules, which narrow down the potential asset list. For example, for the crypto family, we consider a signal with the partial keyword \texttt{key} as the encryption/decryption key if it is a vectored signal (in most cases, the width starts from 64-bit) and has a ``Data Signal'' related behavioral pattern and the stored value needed for any encrypt/decrypt operation. 
In contrast, in the GPIO IP family, the partial keyword \texttt{data} helps to identify \textit{rdata}, \textit{wdata}, and just \textit{port\_data} or \textit{pad\_data} sometimes. The width for the GPIO port varies from 8-bit to the maximum bit at which the system operates (for instance, 32-bit for an ARM Cortex-M0 Microcontroller).
These all have \texttt{data} related behavioral patterns and store important values that are read/written to a register. 
In both cases, \texttt{key} and \texttt{data} can store potentially secret information that must be protected. Still, the detection and classification methods differ based on partial keyword group, width, signal types, and signal behavioral patterns. 
The filtered list is the set of candidate potential assets. 

\subsubsection{Refinement Stage\label{sssec:RS}}
Here, the tool identifies the ``root'' (original sources) of a candidate asset from the Input and Output ports of the TOP module. Any candidate potential asset that is related to the I/O Ports of the TOP module(for a complete IP) or in an important module (where the TOP module is not present) can be considered as a Potential Primary Asset.
At the beginning of this stage, the tool checks a candidate potential asset for its type (Input, Output, Net), behavioral patterns, and module information. Several situations can occur, as follows. 

\paragraph*{Case 1} If the candidate belongs to the Input or Output port of the TOP module, the tool appends it to the potential primary asset list.

\paragraph*{Case 2} If it does not belong to the TOP module, the tool finds the candidate's connection with the Input and output ports of the TOP module by traversing through the instantiations throughout the IP. Then, the tool appends the connected port from the TOP module to the potential primary asset list. If the tool does not find any interconnection between the candidate and the I/O ports of the TOP module, (1) the tool ignores the candidate (maybe a Secondary asset) when the candidate-containing module is instantiated inside the TOP module directly or indirectly through another submodule, or (2) the tool considers the candidate a primary asset when the candidate-containing module is not instantiated inside the TOP module directly or indirectly through another submodule.

\paragraph*{Case 3} If the candidate is a Net-type signal, then the tool identifies the assignments and connections through the instantiations related to the module to which the candidate belongs. The tool detects the candidate's connection(if any) with I/O ports. If the tool can detect any connection between the candidates and I/O ports throughout the IP, the tool repeats \textit{Case 1} and \textit{Case 2}.

After completing the process for each candidate, the tool removes duplicate potential primary assets that can be added multiple times due to the relationship with multiple candidates from multiple modules.
After this stage, the tool lists the final potential primary assets for an \ac{IP} as output.

\section{Experimental Work\label{sec:experiments}}

\subsection{Experimental Setup}

To validate our proposed approach, we implemented a proof-of-concept tool using Python 3.11. 
The tool processes RTL source files, such as those from opencores.org and GitHub. 
In this version, we target Verilog and SystemVerilog files. 
The dataset we use for evaluation is described in~\autoref{tab:IP-information}.
Overall, the asset search space (i.e., the number of signals an engineer would need to consider in manual analysis) comprises 9875 signals. 
Our tool requires the RTL source file directory and Top Module directory as inputs and operates through five stages as described in~\autoref{sec:proposed}. 
The tool's effectiveness is demonstrated by its application to various IP families, producing a detailed list of potential primary assets for each IP in each family. 
We evaluated the accuracy of asset detection by comparing the tool's output against our manually identified asset list. Our list consists of assets defined by the designers in documentation and comments and manually selected assets by a panel of people with experience in Verilog design and security analysis.
In this experiment, we did not consider ``Clock" and ``Reset" signals.

\begin{table}[t!]
\centering
\caption{Characteristics of the IP families used in evaluation}
\label{tab:IP-information}
\begin{tabular}{@{}cccc@{}}
\toprule
IP Family            & No. of IPs & No. of Files & No. of Lines \\ \midrule
Crypto               & 22            & 190          & 61934        \\
Interface-GPIO       & 12            & 16           & 5809         \\
Interface-Peripheral & 13            & 34           & 5118         \\ \bottomrule
\end{tabular}
\end{table}

\subsection{Results}

\begin{figure}[t]
\centering
\subfloat[Crypto]{\label{fig:Crypto}\includegraphics[width=0.30\columnwidth]{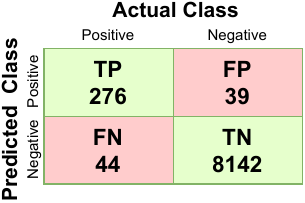}}\hfill
\subfloat[Interface-GPIO]{\label{fig:GPIO}\includegraphics[width=0.30\columnwidth]{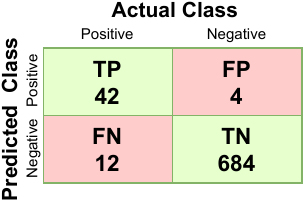}}\hfill
\subfloat[Interface-Peripheral]{\label{fig:Peripheral}\includegraphics[width=0.30\columnwidth]{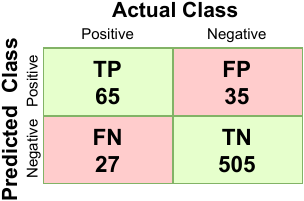}}
\caption{The confusion matrix for three IP families.\label{fig:confusion_matrix}
}
\end{figure}

Our tool's Accuracy and F1 score (out of 1) for the crypto family was 0.9902 and  0.8693, respectively. In GPIO, it was 0.9784 and 0.8400, respectively, and for peripherals, it was 0.9019 and 0.6771, respectively. 
Our results show that our proposed approach can achieve a low false positive rate and can significantly reduce the set of potential assets that need to be considered by a designer compared with a purely manual approach, such as that suggested by the IEEE White Paper~\cite{ieee_p3164_working_group_asset_2024}.
Notably, when we tried the GPIO example from~\cite{ieee_p3164_working_group_asset_2024}, our tool successfully identified the corresponding assets. 
We also evaluated our tool's identified assets against our list of manually-identified assets. 
The results are presented in~\autoref{fig:confusion_matrix} as confusion matrices that show the true positives (TP), true negatives (TN), false positives (FP), and false negatives (FN), where the ``positive'' class is that the signal is an asset. 

When the tool identified false positives and showed inaccuracies, we found these were often due to several reasons, including atypical signal names, incorrect spelling, improper spacing, and atypical abbreviated signals. 
Using external packages for nets and port declaration in RTL files and multi-level instantiations in large IP blocks also reduced accuracy.

The results highlight the potential of our proposed approach as a starting point for automated asset identification which can be used to support downstream activities such as automatic CWE scanning and creating properties for formal verification.

\subsection{Discussion \label{subsection:discussion}}
Asset detection poses challenges due to its potentially subjective nature and time-consuming reliance on manual effort. 
Hence, we were motivated to initiate the development of detection tools as a starting point for ongoing refinement and enhancement of accuracy.
Our results are centered on open-source projects, which might limit its out-of-the-box generalizability to other projects. 
However, our proposed approach remains applicable to 
source code for commercial chips and processors. 
Companies can adopt similar methodologies as discussed in our paper and tailor the developed tool to suit their particular requirements. 

As we observed, most IP designers use ``sensible'' names, often based on organizational styles like in OpenTitan~\cite{opentitan_open_nodate}. 
To use our approach, designers can explore their own styles to help create an initial set of keywords for potential assets and extend our approach to other IP families. 

Our proposed approach can be adapted to different types and sizes of IP families, although the initial adaptation requires some security expertise. 
To adapt the proposed approach for a different IP family or project, practitioners should use the initial manual process we presented to identify their own keywords or unique behaviors. Afterward, they can apply the automated approach to other related design projects. 

To prepare the keyword list, one can extract the input and output port names and net names from different projects of an IP family. 
Our results on open-source designs show that, despite different designers, it appears that patterns in names and behaviors exist, possibly due to design guidelines and conventions. 
Our approach of identifying ``partial keywords'' (as described in~\autoref{ssec:IED}) can help capture potential asset candidates.  
For a new project or design style, important signals need to be sorted into different categories of signal types (e.g., control signals, configuration signals, as in~\autoref{ssec:BP}).
Each category is linked to confidentiality, availability, integrity, or undermined behavior properties, as in~\cite{ieee_p3164_working_group_asset_2024}, so an identified asset's potential relevance to these properties can be reported. 
Using the sets of partial keyword groups and signal categories, all classification rules can be formulated for an IP class using the following strategy as follows. 

If a known signal, e.g., \texttt{test\_signal}, with one or more security objectives, according to to~\cite{ieee_p3164_working_group_asset_2024}, belongs to a partial keyword group, for instance, $PKG_{x}$, and has a \texttt{Control\_Signals[]} pattern, any signal that belongs to the same partial keyword group, $PKG_{x}$ and \texttt{Control\_Signals[]} pattern list, can be identified as a \textbf{Potential Asset}. A Signal, $s$, belongs to the known keyword group, $PKG_{x}$, and has \texttt{Control\_Signals[]} properties but not an Input signal, maybe a net or variable signal. In that case, to find out the potential \textbf{Primary Asset}, we need to identify the blocking/non-blocking assignment or continuous assignment where the signal, $s$, is assigned with a single-bit input port on the left-hand side. The Refinement Stage described in \autoref{sssec:RS} completes this process. 

\section{Conclusions and Future Work\label{sec:conclusion}}
In this paper, we proposed an automated approach for identifying potential assets in Verilog designs based on observations that we made from analyzing open-source hardware designs. 
Identified assets can then be used in downstream security-related tasks. 
Moving forward, our future work will refine accuracy by incorporating additional behavioral abstraction detection to associate identified assets with confidentiality, integrity, and availability attributes, which can provided as collateral to IP consumers. 
This will also incorporate structural analysis in the source code. 
Additionally, we plan to use identified assets to map IPs to potentially relevant \acp{CWE} to help with verification.

\IEEEtriggercmd{\balance}
\IEEEtriggeratref{12}

\bibliographystyle{IEEEtran}
\bibliography{IEEEabrv,trefs}

\end{document}